\documentclass[runningheads]{llncs}
\usepackage[T1]{fontenc}
\usepackage{hyperref}
\usepackage{subcaption}
\usepackage{booktabs,multirow, makecell}
\usepackage{diagbox} 
\usepackage{xcolor}
\usepackage{graphicx}
\usepackage{enumitem}
\usepackage{stfloats}
\usepackage{amsmath,amssymb}
\usepackage{diagbox}
\usepackage{algorithm}
\usepackage{setspace}
\usepackage[noend]{algpseudocode} 
\usepackage{float}  
\setlist[itemize]{leftmargin=*}
\usepackage{pifont}
\usepackage{amssymb}
\usepackage[format=plain,textfont=footnotesize,labelfont={bf,sf, footnotesize},tableposition=top,justification=justified]{caption}

\newcommand{\Nash}{\mathrm{Nash}}
\newcommand{\algacro}{PGTS}
\newcommand{\subgame}{sub-team game}
\newcommand{\subgameintro}{sub-game}

\newcommand{\team}{\ast}
\renewcommand{\team}{\bullet}
\renewcommand{\team}{\ast}

\newcommand{\w}{\text{v}}
\let\oldmarginpar\marginpar
\renewcommand{\marginpar}[1]{\oldmarginpar{\color{blue}\footnotesize #1}}

\begin{document}
\title{Offline Nash Solvers Meet Online Tree Search in Multi-Agent Games on Graphs}
\titlerunning{Online Tree Search with Offline Nash Solvers}
%
\author{
Mukesh Kumar\thanks{Corresponding author.} 
\and
Yue Guan \and
Panagiotis Tsiotras
}

\institute{
School of Aerospace Engineering, Georgia Institute of Technology, Atlanta, GA, USA\\
\email{\{mkumar361,yguan44,tsiotras\}@gatech.edu}
}
\authorrunning{M. Kumar et al.}
\maketitle              
\begin{abstract}
Computing Nash equilibrium policies in multi-agent Pursuit-Evasion games (PEG) is challenging due to the exponential growth of the joint state and action spaces with the number of agents. 
Existing approaches either rely on offline equilibrium approximations, which may lack adaptability during execution, or online planning methods, which suffer from large branching factors. 
In this work, we propose Primitive-Guided Tree Search (\algacro{}), a hybrid framework that integrates offline exact Nash equilibrium computation with online tree search:
\algacro{} first solves a collection of smaller, tractable \subgameintro{}s offline;
at deployment, \algacro{} performs online tree search at each time step, using the optimal \subgameintro{} policies and value functions to guide tree expansion and estimate leaf-node values.
Extensive experiments on varied graph topologies, including real-world networks, demonstrate that \algacro{} significantly outperforms state-of-the-art learning and heuristic baselines, while maintaining robust performance against adversaries.

\end{abstract}


\section{Introduction}

Multi-agent decision-making under adversarial conditions arises in a wide range of applications, including monitoring and surveillance, urban security, border patrol, and autonomous pursuit tasks~\cite{Pursuit_survey_paper,urban_network_security}. 
A canonical model for studying such interactions in discrete settings is the graph-based \emph{Pursuit--Evasion} game (PEG) \cite{guan2021bounded,parsons2006pursuit}. 
%
%
The standard solution concept for such games is the Nash equilibrium. While the existence of equilibrium policies is well established \cite{fink_paper}, computing equilibria exactly in realistic multi-agent settings remains a significant challenge. 
As the number of agents increases, both the state space and the joint action space grow rapidly. 
This combinatorial growth renders exact equilibrium computation impractical even for moderately sized environments, motivating the development of scalable approximation and planning techniques.


Existing approaches to approximating solutions in multi-agent settings can be broadly classified into three categories: offline, online, and hybrid. 
Offline approaches, such as Policy Space Response Oracles (PSRO)~\cite{psro_survey} and its variants, provide a general framework for computing approximate equilibria by iteratively training best-response policies and maintaining a growing meta-game over the policy space.
In the Pursuit-Evasion context, MT-PSRO~\cite{pretrain_PSRO} combines PSRO with multi-task pretraining, and uses MAPPO~\cite{mappo} as the best-response oracle. 
While offline methods enable fast decision-making during execution, their training complexity can grow quickly with the number of agents.
Moreover, the agents' performance can degrade significantly when they encounter adversarial behaviors that differ from those encountered during training.

Online planning methods, such as Monte Carlo Tree Search (MCTS)~\cite{Mcts-survey}, adapt to changing situations and reason about strategic interactions at execution time by simulating a large number of future trajectories.
Although originally introduced to solve sequential in-turn games, 
extensions of MCTS to simultaneous-move games have been explored through Simultaneous-Move MCTS (SM-MCTS)~\cite{mctssmmove}.
Unlike offline methods, online search methods can reason directly about the current game state, but their decision quality depends critically on the ability of the search process to identify and explore promising future trajectories.
%
In multi-agent games, this exploration becomes increasingly difficult because the number of joint actions, and hence the branching factor, grows exponentially with the number of agents.
As a result, practical implementations often rely on carefully designed heuristics, or domain-specific knowledge to focus the search on promising regions of the action space~\cite{heuristic_mcts_2,nijssen2013monte,heuristic_mcts}.

More recently, hybrid approaches have sought to combine the strengths of offline computation and online planning. 
Methods such as AlphaZero and its variants \cite{simultaneousalphazero,schmid2023student,alphazero} have demonstrated that offline-computed priors can significantly improve the efficiency of online search.  
These methods focus primarily on two-player games, although similar ideas have also been explored in multi-agent settings. 
For example, NSGZero \cite{NSGZero} combines neural policy and value function approximation with MCTS to learn Pursuer policies in network security games on graphs.
%
%
While these approaches achieve strong empirical performance, they rely on learning accurate policy and value approximations, often requiring a substantial amount of training data. 
 
\begin{figure*}[t]
    \centering
    \includegraphics[width=0.99\linewidth]{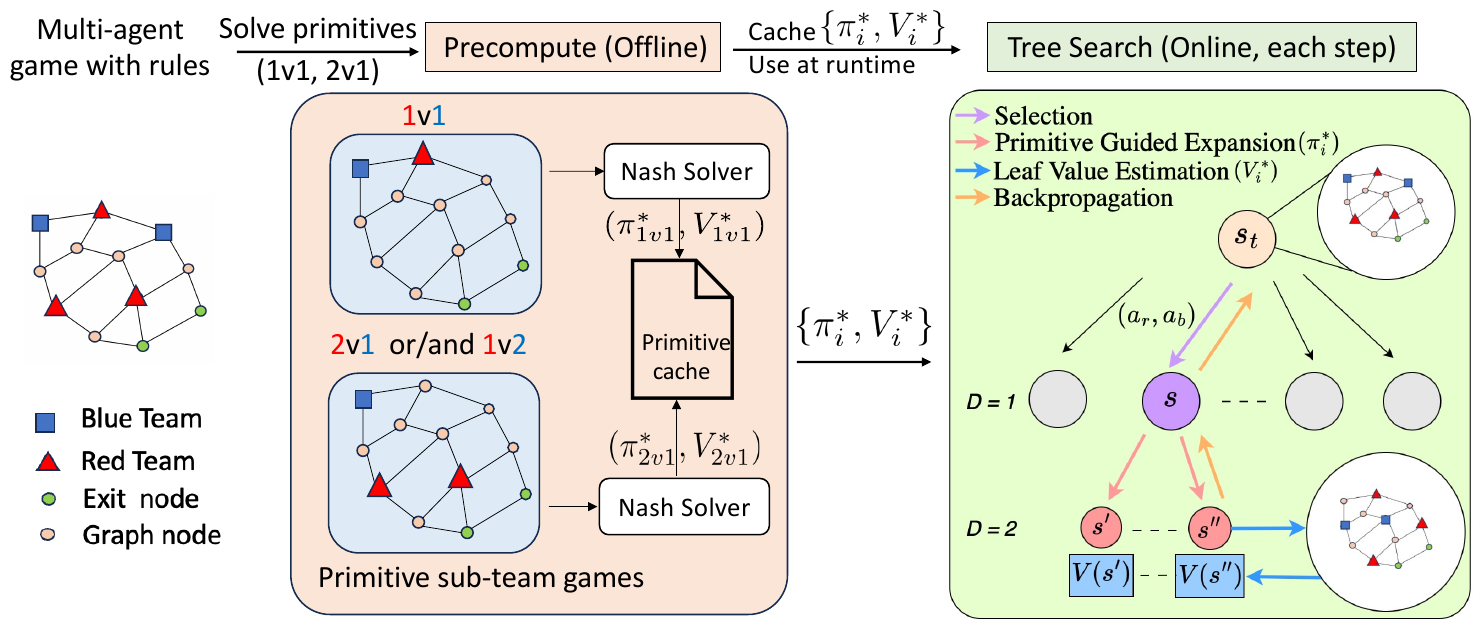}
        \vspace{-0.15in}
        \caption{Illustration of the hybrid approach for Pursuit-Evasion games. 
        \textbf{Left}: a game configuration with three Pursuers (Red), two Evaders (Blue), and two exit nodes (Green). 
        \textbf{Right}: pipeline of the proposed \algacro{} algorithm. 
        Primitive sub-team games (e.g., $1\w1$, $2\w1$) are solved offline to obtain equilibrium policies and value functions. 
        Online tree search iterates among four steps: \texttt{Selection, Expansion}, \texttt{Leaf Value Estimation}, and \texttt{Backpropagation}.
        The primitives computed offline are utilized to guide roll-out during \texttt{Expansion} and to perform \texttt{Leaf Value Estimation}. 
        }
        \vspace{-0.1in}
        \label{proposedsolution}

\end{figure*}

Motivated by the strengths and limitations of existing approaches, we propose Primitive-Guided Tree Search (\algacro{}), a hybrid framework that combines offline game-theoretic computation with online planning.
Rather than computing policy and value functions for the original multi-agent game directly, the offline phase of \algacro{} decomposes the original game into a collection of smaller, computationally tractable \subgameintro{}s called \textit{primitive \subgame{}s}, each involving only a subset of agents, and solves exactly these primitive \subgame{}s via value iteration.
At deployment, \algacro{} performs online tree search and leverages the offline-computed solutions.
%
Specifically, \algacro{} introduces two adaptations that make tree search more efficient and effective than existing variants.
First, the equilibrium policies of the primitive \subgame{}s are used to guide \emph{joint-action} rollouts, allowing the search to focus its computational effort on the most relevant regions of the joint action space.
Second, the value functions of the primitive sub-games are used to approximate the leaf-node values of the MCTS search tree, allowing the search to terminate at a shallower depth while still maintaining good estimates of future game outcomes.
Together, these adaptations keep both the branching factor and the effective search depth manageable, while preserving strong empirical performance.
At the same time, since \algacro{} rolls out actions in the full joint action space, it retains the ability to reason about coordinated policies among all agents, thereby compensating for the lack of team-level coordination of the primitive games, which involve only a subset of the agents.
Figure \ref{proposedsolution} illustrates the core idea of the hybrid algorithm for a multiple-agent Pursuit-Evasion game.

The remainder of the paper is organized as follows.
Section~\ref{sec:formulation} presents the formulation of the multi-agent Pursuit-Evasion game (PEG) on a graph, introduces the notion of Nash equilibrium, and presents the corresponding value-iteration algorithm for its computation.
Section~\ref{sec:game-decomp} presents the decomposition of the full game into primitive games involving subsets of agents.
Section~\ref{sec:tree-search} develops the \algacro{} algorithm 
while
Section~\ref{sec:experiment} provides a comprehensive empirical analysis of \algacro{} under different PEG configurations, compares it with baseline algorithms, and presents ablation studies to demonstrate the contribution of each component.
Finally, Section~\ref{sec:conclusion} concludes the paper with a summary of possible future extensions.

\section{Problem Formulation}
\label{sec:formulation}

We consider a graph-based multi-agent Pursuit--Evasion game between a Red team of Pursuers and a Blue team of Evaders. The game is defined by the tuple $\mathcal{G} = \langle \mathcal{N}, G,V_{\rm e}  ,\mathcal{P}, \mathcal{R}, \Gamma^G, \beta \rangle$, and
the set of agents $\mathcal{N} = \mathcal{N}_r \cup \mathcal{N}_b$ is partitioned into the Red/Pursuers team, $\mathcal{N}_r$, consisting of $n_r = |\mathcal{N}_r|$ agents, and the Blue/Evaders team, $\mathcal{N}_b$, consisting of $n_b = |\mathcal{N}_b|$ agents. 
The underlying environment is a graph $G = (V,E)$, where $V$ denotes the set of vertices and $E \subseteq V \times V$ denotes the set of edges. 
The set $V_{\rm e} \subseteq V$ denotes the set of exit vertices available to the Blue team.
At a time step $t$, the positions of the Red and Blue agents are given by $s_t^r = ( s_t^{r,1}, \ldots, s_t^{r,n_r} )$ and $s_t^b = ( s_t^{b,1}, \ldots, s_t^{b,n_b} )$, respectively.
The joint state is then $s_t = (s_t^r, s_t^b) \in V^{n_r} \times V^{n_b}$, and we denote this joint state space by $\mathcal{S}$. 
For each agent $i \in \mathcal{N}$, the available actions at vertex $s^i \in V$ are given by the set $\mathcal{A}^i(s^i) = \{ v \in V \mid (s^i,v) \in E \} \cup \{ s^i \}$, corresponding to either moving to an adjacent neighbor or staying in place. 
Consequently, for each team $\team \in \{r,b\}$, the feasible action space at joint state $s=(s^r,s^b)$ is given by $\mathcal{A}^{\team}(s^{\team})=\prod_{i\in\mathcal{N}_{\team}}\mathcal{A}^i(s^i)$.
%
%
At each time step, all agents act simultaneously.
Given an action for agent $i$ of the Red team at time $t$, we can define the (Red) team action as $a_t^r = (a_t^{r,1},\ldots,a_t^{r,n_r})$, and similarly for the Blue team, that is,  $a_t^b = (a_t^{b,1},\ldots,a_t^{b,n_b})$.
Given now a joint state $s_t$ and a joint action $a_t=(a_t^r,a_t^b)$, the game transitions deterministically to the next state according to the transition function $s_{t+1} = \mathcal{P}(s_t,a_t)$. 
While, for the sake of simplicity, we focus on deterministic transitions in this work, stochastic transition kernels can be incorporated with only minor modifications within our approach.
Furthermore, we assume a fully observable environment in which both teams have access to the complete joint state and full knowledge of the game dynamics.

The inter-team interactions are specified by the game rules $\Gamma^G$. 
Specifically, $\Gamma^G$ defines the capture, escape, post-capture, and terminal conditions.
In our setting, a Blue agent is captured whenever it lies within a capture radius of $r_c$ hops from at least one Red agent.
%
Similarly, a Blue agent escapes whenever it is within an escape radius of $r_e$ hops from an exit node. 
The reward function $\mathcal{R}$ is defined at the team level,
and is assumed to be zero-sum, with the Red team maximizing and the Blue team minimizing the expected cumulative reward, with $\beta \in (0, 1)$ serving as the discount factor; see equation~\eqref{eqn1} below.
At any time step, the Red team receives a reward of +1 whenever a Blue agent is captured and a reward of -1 whenever a Blue agent escapes. The Blue team receives the corresponding opposite reward.  
The game terminates when all Blue agents have either been captured or escaped, or when another terminal condition specified by $\Gamma^G$ is satisfied.
%


The game rule set $\Gamma^G$ is sufficiently general to accommodate both finite and infinite-horizon formulations.
To highlight the computational challenges inherent in solving the resulting multi-agent game $\mathcal{G}$, we focus the remainder of the discussion on the infinite-horizon setting, which provides a convenient framework for illustrating the scalability issues that motivate our approach.

\subsection{Nash Equilibrium}

We consider a Markovian policy $\pi^\team$ for each team 
$\team \in \{ r,b\}$, which is a mapping  $\pi^\team:\mathcal{S} \xrightarrow{} \Delta(\mathcal{A}^\team(s^\team))$  from the joint state space to a distribution over valid moves at that team state. 
For notational convenience, we will also often write $\pi^\team (a^\team |s)$ to denote the probability assigned to team action $a^\team\in\mathcal{A}^\team(s^\team)$ at joint state $s$.
At any joint state $s$, the Red team selects an action $a^r \in \mathcal{A}^r(s^r)$, 
and the Blue team selects an action $a^b \in \mathcal{A}^b(s^b)$.
Given a pair of policies $(\pi^{r}, \pi^{b})$ for the two teams, the induced value for each state $s  \in \mathcal{S}$ is defined as
\begin{equation} \label{eqn1}
\mathcal{V}^{\pi^{r}, \pi^{b}}(s) = \mathbb{E} \left[ \sum_{t=0}^{\infty} \beta^t \mathcal{R}(s_t,a^r,a^b) \,\middle|\, a^r \sim\pi^{r},  a^b\sim\pi^{b}, s_0 = s \right]
\end{equation}
where $\beta \in (0,1)$ is the discount factor and $s_t$ denotes the joint state at time $t$.
A Nash equilibrium is then defined as a pair of Markovian policies $(\pi^{r*}, \pi^{b*})$ such that, for all $s \in \mathcal{S}$ and all Markovian
policies $\pi^{r}, \pi^{b}$, the following inequality holds:
\begin{equation} \label{eqsaddle}
\mathcal{V}^{\pi^{r}, \pi^{b*}}(s) \leq \mathcal{V}^{\pi^{r*}, \pi^{b*}}(s) \leq \mathcal{V}^{\pi^{r*}, \pi^{b}}(s),
\quad s \in \mathcal{S}.
\end{equation}

The overall objective is to compute the Nash equilibrium policies $(\pi^{r*}, \pi^{b*})$ that satisfy the saddle-point condition described in \eqref{eqsaddle}.

\subsection{Shapley Value Iteration}
\label{solution_concept}

The standard Shapley value iteration for two-player zero-sum stochastic games~\cite{shapley_book} can be utilized to solve the multi-agent game $\mathcal{G}$ by treating each team as a single decision-maker. 
%
To compute the equilibrium, Shapley value iteration updates are performed to the state-action (Q-value) matrices until convergence. 
Specifically,
let $k \in \{0,1,2,\ldots\}$ denote the iteration index, and  let $\boldsymbol{Q}_k(s) \in \mathbb{R}^{|\mathcal{A}^r(s^r)| \times |\mathcal{A}^b(s^b)|}$ denote the Q-matrix for state $s$ at iteration $k$. 
The update rule for this iterative process is given by
\begin{equation}\label{shapely_operator}
Q_{k+1}(s,a^r,a^b) = \mathcal{R}(s,a^r,a^b) + \beta \,
\Nash(\boldsymbol{Q}_k(s')),
\end{equation}  
where 
$\mathcal{R}(s,a^r,a^b)$ is the immediate reward to the Red team and $s' = \mathcal{P}(s,(a^r,a^b))$ is next state after applying the action pair $(a^r,a^b)$.
%
The value $\Nash(\boldsymbol{Q}_k(s^{'}))$ denotes the Nash equilibrium payoff of the stage game defined by the matrix $\boldsymbol{Q}_k(s^{'})$. 
From the maximizing Red team’s perspective, the Nash value at any joint state $s$ corresponds to the max-min value
\begin{align}
    \Nash(\boldsymbol{Q}(s)) = \max_{\pi^r(s) \in \Delta(A^r(s^r))} \min_{a^b} \; \sum_{a^r} \pi^r(a^r|s) Q(s, a^r, a^b),
\end{align}
which can be obtained by solving the following linear program: 
%
\begin{align}\label{nashsolver}
\max_{v,\;\pi^r(s)}  & v \nonumber\\
\text{s.t.} \quad & v\,\mathbf{1}-\pi^r(s)^\top\boldsymbol{Q}(s) \;\le\; 0 , \\
& \mathbf{1}^\top \pi^r(s) \;=\; 1, \quad \pi^r(s) \;\ge\; 0, \nonumber
\end{align}  
where $\pi^r(s) \in \Delta (\mathcal{A}^r(s^r))$ 
is the  Red team's 
policy and $\mathbf{1}$ is the vector of ones of compatible dimension.
The solution $v^*$ of \eqref{nashsolver} gives the Nash value $v^* = \Nash(\boldsymbol{Q}(s))$, and $\pi^{r*}(s)$ gives the Nash policy for the Red team at joint state $s$.
The Blue team policy can be obtained from the corresponding dual program or directly from the dual variables of~\eqref{nashsolver}; see~\cite{shapley_book}.
Under standard assumptions~\cite{shapley_book}, the Shapley operator~\eqref{shapely_operator} is a contraction and the sequence $\{\boldsymbol{Q}_k\}$ converges to a unique fixed point $\boldsymbol{Q}^{*}$. 
For every state $s \in \mathcal{S}$, the corresponding optimal Nash value is then given by 
$V^{*}(s) = \Nash(\boldsymbol{Q}^{*}(s))$, and the equilibrium policy pair $\pi^* = (\pi^{r*},\pi^{b*})$ is obtained by solving for the Nash equilibrium of the stage game induced by $\boldsymbol{Q}^{*}(s)$.

While Shapley’s value iteration provides a principled solution method for two-player zero-sum stochastic games, each value update requires computing the Nash value of a local matrix game at each state.
This makes a na\"{\i}ve adoption of Shapley’s value iteration to two-\emph{team} zero-sum stochastic games computationally prohibitive.
In particular, both the number of linear programs solved at each iteration and the size of the corresponding Q-function grow exponentially with the number of agents.
As a result, exact computation quickly becomes intractable even for moderately sized teams and environments.

To address this challenge, we exploit the observation that many large-scale Pursuit--Evasion games are composed of recurring local interactions involving only a small number of agents. 
Rather than solving the full game directly, we first solve a collection of smaller representative games, referred to as \emph{primitive \subgame{}s}, and use their equilibrium solutions as reusable building blocks. 
These \textit{primitive \subgame{}s} involve only a small number of agents from each team (e.g., 1 Red vs. 1 Blue, 2 Red vs. 1 Blue, etc) and are played on the same graph with identical engagement rules
$\Gamma^G$. 
Since these primitive games contain only a few agents, their exact Nash policies and value functions can be computed efficiently offline and reused during online execution.

A natural use of these primitive solutions is through a \emph{sub-team decomposition}~\cite{matching1,Daigospaper}
where the original game is decomposed into a collection of primitive interactions and the resulting equilibrium policies are used directly for decision making. 
While computationally efficient, such an approach treats the primitive games independently and therefore captures only limited coordination across different groups of agents. 
To overcome this limitation, we propose \emph{Primitive-Guided Tree Search} (\algacro{}), which incorporates the pre-computed primitive solutions into an online tree-search framework. 
The primitive policies guide exploration, while the primitive value functions provide estimates of future game outcomes. 
As a result, \algacro{} enables coordinated planning over the full agent team while preserving the computational advantages of the offline-computed primitive games.

We now describe the construction of these {primitive \subgame{}s} and the computation of their corresponding Nash equilibrium policies and value functions.
To operationalize the offline phase, we compute exact Nash equilibrium solutions for two {primitive \subgame{}s}: 1 Red vs.\ 1 Blue ($1\w1$) and 2 Red vs.\ 1 Blue ($2\w1$);
the approach is similar for subgames with more agents.
Unlike the original multi-agent game, the restricted joint state spaces of the two subgames, namely, $\mathcal{S}_{1\w1} = V \times V$ and $\mathcal{S}_{2\w1} = V^2 \times V$, permit exact policy computation over the entire state space. 
Using the solver from \cite{guan2021learning}, we derive the optimal value functions $V^{*}_{1\w1}: \mathcal{S}_{1\w1} \to \mathbb{R}$ and $V^{*}_{2\w1}: \mathcal{S}_{2\w1} \to \mathbb{R}$, along with the corresponding Nash policy pairs $\pi^{*}_{1\w1} = (\pi^{r*}_{1\w1}, \pi^{b*}_{1\w1})$ and $\pi^{*}_{2\w1} = (\pi^{r*}_{2\w1}, \pi^{b*}_{2\w1})$. 
These policies map primitive states to probability simplices over joint actions. 
For instance, at state $s_{1\w1} = (s^{r}, s^{b}) \in \mathcal{S}_{1\w1}$, the policies $\pi^{r*}_{1\w1}(s_{1\w1}) \in \Delta(\mathcal{A}^{r}(s^r))$ and $\pi^{b*}_{1\w1}(s_{1\w1}) \in \Delta(\mathcal{A}^{b}(s^b))$ return the optimal mixed policy for the Red and Blue agents, respectively, with $2\w1$ policies scaling analogously over $\mathcal{S}_{2\w1}$. 
We first describe the sub-team decomposition in the next section, which also serves as a baseline for our proposed hybrid approach.
\section{Adaptive Sub-Team Decomposition} \label{sec:game-decomp}

In this section, we formalize a decomposition-based approach that utilizes primitive \subgame{}s to approximate the multi-agent interaction of the original game.
We describe the approach from the Red team's perspective, which acts as the maximizer.

A key observation underlying the decomposition approach is that the offline-computed primitive \subgame{} value functions quantify the equilibrium outcome that can be expected from the corresponding local interactions. 
Therefore, by partitioning the multi-agent game into a collection of primitive \subgame{}s and aggregating their values, we can obtain a surrogate measure of the evaluation of the joint state. This surrogate can then be used to identify a favorable assignment of agents to primitive interactions and induce a localized \subgame{} policy. 
To formalize the proposed decomposition approach, let $s_t = (s^r_t, s^b_t)$ denote the joint state of the game at time $t$.
Given $\mathcal{R} \subseteq \mathcal{N}_r$ and $\mathcal{B} \subseteq \mathcal{N}_b$, We first define a  sub-team game as a  tuple $g = \langle (\mathcal{R}, \mathcal{B}), G, V_{\text{e}}, \mathcal{P}, \mathcal{R}, \Gamma^G, \beta \rangle$, isolating a subset of Red agents $\mathcal{R}$ and Blue agents $\mathcal{B}$ into a localized interaction while inheriting the underlying structure from the original game $\mathcal{G}$. 
Each sub-team game $g$ corresponds to one of the available primitive types, meaning the team sizes are restricted to $(|\mathcal{R}|, |\mathcal{B}|) \in \{(1,1), (2,1)\}$ for our illustrative case.
For a given sub-team game $g$, the local state is defined as $s^{g}_t = (s^{\mathcal{R}}_t, s^{\mathcal{B}}_t)$, where $s^{\mathcal{R}}_t$ and $s^{\mathcal{B}}_t$ denote the states of the agents present in $\mathcal{R}$ and $\mathcal{B}$, respectively.

\begin{figure}[t]
    \centering
    \includegraphics[width=0.90\linewidth]{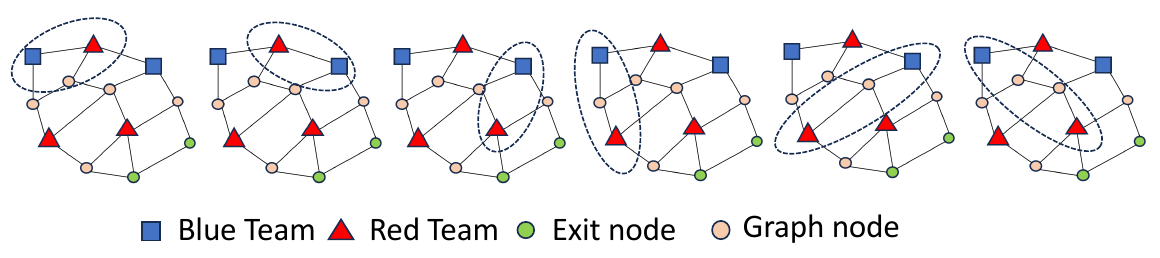}
    \vspace{-0.3cm}
    \caption{Illustration of all six valid decompositions for a configuration with 3 Red agents and 2 Blue agents into primitive \subgame{}s on a random graph with two goal nodes. Each decomposition consists of one $1$v$1$ \subgame{} (indicated by the dotted ellipsoid) and one $2$v$1$ \subgame{}, formed by pairing the remaining Blue agent with the other two Red agents.}
    \label{decompositions}
\vspace{-0.5cm}

\end{figure}

We define a decomposition $d(s_t)$ of the joint state $s_t$ at time $t$ as a collection of sub-team games, i.e., $d(s_t) = \{ g_1, g_2, \dots, g_L \},$ where $L$ denotes the total number of sub-team games.
Note that $g_1, g_2, \dots, g_L$ depend on $t$, but we drop this dependence for notational convenience.
%
%
The set of admissible decompositions, denoted by $\mathcal{D}$, consists of all such collections satisfying the partition constraints
\begin{equation}\
    \bigcup_{\ell=1}^L \mathcal{R}_\ell = \mathcal{N}_r, \quad \bigcup_{\ell=1}^L\mathcal{B}_\ell = \mathcal{N}_b, \quad \text{with} \quad \mathcal{R}_\ell \cap \mathcal{R}_{\ell'} = \mathcal{B}_\ell \cap \mathcal{B}_{\ell'} = \emptyset \quad \forall \ell \neq \ell'.
\end{equation}
That is, each agent is assigned to exactly one \subgame{}, and no agent participates in more than one \subgame{}. 
For example, consider a configuration with three Red agents and two Blue agents. 
Utilizing the $1\w1$ and $2\w1$ primitive \subgame{}, there are six valid decompositions as illustrated in Fig.~\ref{decompositions}.

Each decomposition induces a distinct set of local interactions and can lead to  different outcomes; the goal is therefore to identify the assignment that best approximates the outcome of the full game.
Given a decomposition $d\in \mathcal{D}$, we use a surrogate value function to evaluate the decomposition  at $s_t$ by summing the pre-computed value functions of its constituent sub-team games, i.e.,
\begin{equation}
    \bar{V}(d, s_t) = \sum_{g_\ell \in d(s_t)} V^*_{g_\ell}(s^{g_\ell}_t),
\end{equation}
where $V^*_{g_\ell}$ is the corresponding primitive sub-team game value function ($V^*_{1\w1}$ or $V^*_{2\w1}$ based on the team sizes $|\mathcal{R}_\ell|, |\mathcal{B}_\ell|$).
The optimal decomposition is the partition that maximizes this surrogate value:
\begin{equation}
    d^*(s_t) = \arg\max_{d \in \mathcal{D}} \bar{V}(d, s_t).
\end{equation}

Once the optimal decomposition $d^*(s_t)$ is computed, it induces a decentralized \textit{sub-team policy}. 
Under this policy, action selection is executed at the sub-team game level. 
For each assigned \subgame{} $g_\ell \in d^*(s_t)$, 
the subset of Red team agents $\mathcal{R}_\ell$ collectively observes the local joint state $s^{g_\ell}_t$ and samples a joint sub-team action $a^{\mathcal{R}_\ell} \sim \pi^{r*}_{g_\ell}(s^{g_\ell}_t)$ according to the pre-computed primitive sub-team Nash policy ($\pi^{r*}_{1\w1}$ or $\pi^{r*}_{2\w1}$ based on the team sizes $|\mathcal{R}_\ell|, |\mathcal{B}_\ell|$).
Each individual agent $i \in \mathcal{R}_\ell$ then executes its corresponding local action $a^i$ derived from the sampled joint action $a^{\mathcal{R}_\ell}$. 
This decomposition, evaluation, and assignment process is executed dynamically at every time step throughout the game's execution.
However, as discussed previously, a decomposition-based approach has limited ability to capture coordination at the team level. 
This limitation motivates the tree-search algorithm presented in the next section, which explicitly accounts for coordination among agents at the team level by sampling joint actions via rollouts.
\vspace{-0.1in}
\section{Primitive-Guided Tree Search}
\label{sec:tree-search}

In this section, we present \algacro{}, a hybrid online planning framework for multi-agent games that combines offline-computed primitive sub-team game solutions with online tree search,
building on the Simultaneous-Move Monte Carlo Tree Search (SM-MCTS) framework \cite{mctssmmove}. 
As discussed already, directly applying tree search in the joint action space is computationally infeasible due to the exponential branching factor. 
Our approach addresses this challenge by incorporating structure from the primitive \subgame{}s into both expansion and leaf node value estimation. 

We describe the algorithm from the Red team's perspective (maximizer); the tree is built in exactly the same way for the Blue team (minimizer), with the only difference being the selection and backpropagation steps. 
At each execution time step $t$, the current joint state $s_t = (s_t^r, s_t^b)$ initializes the root node of the tree $\mathcal{T}$. 
The tree is built incrementally through $M$ repeated simulations where $M$ denotes the total simulation budget before a final action is executed in the environment. 
Each simulation executes the following phases (Fig.~\ref{proposedsolution}):
\begin{enumerate}
    \item \texttt{Selection:} The tree is traversed by selecting joint actions until a terminal or unexpanded node is reached. 
    \item \texttt{Expansion:} The unexpanded node is expanded to generate new child nodes.
    \item \texttt{Leaf Value Evaluation:} The newly generated nodes are evaluated to estimate the future game outcome.
    \item \texttt{Backpropagation:} The resulting reward or surrogate estimate is passed back up the visited trajectory to update the node statistics.
\end{enumerate}
Since exhaustive search is computationally infeasible in large multi-agent games, planning is performed over a finite horizon by limiting the maximum search depth to $D$. Our primary contributions lie in designing the structured \texttt{Expansion} phase and a scalable \texttt{Leaf Value Evaluation} mechanism tailored to the multi-agent setting. The pseudo-code for the overall approach is provided in Algorithm~\ref{treesearch-algo}.

\subsection{Primitive-Guided Expansion} 

A key challenge in multi-agent tree search is to identify promising actions without exhaustively expanding the combinatorially large joint action space. While heuristic action-sampling schemes can reduce the branching factor, they often fail to exploit the strategic structure of the underlying game. To address this challenge, we perform a \textit{primitive-guided expansion} to guide expansion toward actions that are likely to be relevant under strategic play, thereby focusing the search budget on strategically promising regions of the action space.

During the expansion of a node $n$ in the search tree $\mathcal{T}$, let $s_n =(s_n^r,s_n^b)$ denote the  joint state associated with node $n$. Each Blue agent $j \in \mathcal{N}_b$ identifies its $k_r$ nearest Red opponents, denoted by the neighborhood set $\mathcal{N}_{k_r}(j) \subseteq \mathcal{N}_r$. We then sample localized interactions using the pre-computed primitive \subgame{} Nash policies:

\begin{itemize}
    \item \textbf{1v1 samples:} $(a^{r,i}, a^{b,j}) \sim \pi^{*}_{1\w1}( s^{i}_{n}, s^{j}_{n})$ for $i \in \mathcal{N}_{k_r}(j)$.
    Specifically, for each Blue agent $j$, we enumerate individual Red agents in its neighborhood set $\mathcal{N}_{k_r}(j)$. 
    For each pair of a Red agent and a Blue agent, we sample their joint actions according to $\pi^{*}_{1\w1}$ based on their current joint state.
    \item \textbf{2\w1 samples:} $(a^{r,i_1}, a^{r,i_2}, a^{b,j}) \sim \pi^{*}_{2\w1}((s^{i_1}_{n}, s^{i_2}_{n}), s^{j}_{n})$ for unordered pairs 
    $\{i_1,i_2\} \subseteq \mathcal{N}_{k_r}(j)$.
    Similar to the 1\w1 case, we enumerate pairs of two Red agents both in Blue agent $j$'s neighborhood $\mathcal{N}_{k_r}(j)$.
    For each two-Red–one-Blue group, we sample the three agents' actions from $\pi^{*}_{2\w1}$.
\end{itemize}
We aggregate these local action samples across all Blue agent neighborhoods to construct a restricted candidate action set for the Red team $\tilde{\mathcal{A}}^r(s_{n}^r) \subseteq \mathcal{A}^r(s_{n}^r)$
and the Blue team $\tilde{\mathcal{A}}^b(s_{n}^b) \subseteq \mathcal{A}^b(s_{n}^b)$.
%
The restricted joint action space evaluated at node $n$ is therefore:
$\tilde{\mathcal{A}}(s_{n}) = \tilde{\mathcal{A}}^r(s_{n}^r) \times \tilde{\mathcal{A}}^b(s_{n}^b) \subseteq \mathcal{A}(s_{n})$. 
To maintain exploration and prevent the search from overcommitting to the primitive priors, we augment the restricted action set with heuristic shortest-path actions toward relevant opponents/exits, as well as randomly sampled joint team actions.
Node $n$ is then expanded by instantiating a new child node $n'$  for each joint action $a = (a^r,a^b) \in \tilde{\mathcal{A}}(s_{n})$.
%
%
%
This structured expansion mechanism reduces the branching factor  
while biasing the search toward strategically relevant actions.
Additionally, to maintain tractability, agents that are captured or have escaped upon reaching the new state $s_{n'}$ are removed based on the game rules $\Gamma^G$, and only the remaining agents are simulated thereafter.

\subsection{Leaf Value Estimation via Sub-team Surrogates}
\label{leaf value estimation}

The tree-search algorithm requires an estimate of the value for each leaf node in the search tree $\mathcal{T}$.
Rather than performing expensive Monte-Carlo rollouts or learning a separate value function, we approximate the value of a leaf node using the decomposition framework introduced in Section \ref{sec:game-decomp}, which leverages the pre-computed value functions of the \textit{primitive \subgame{}s}.

Let $s$ denote the joint state, getting an estimate of the value of state $s$
requires solving $\max_{d\in\mathcal D}\bar V(d,s)$ over the set of admissible decompositions $\mathcal{D}$.
However, finding the optimal decomposition scales combinatorially with the number of agents.
To obtain a scalable approximation suitable for online planning, we relax this optimization problem to a structured two-stage sequential matching problem.
    
Crucially, we prioritize extracting the 2v1 interactions first. Higher-order primitives capture essential multi-agent coordination such as joint trapping that isolated 1v1 pairings fundamentally ignore. 
By matching the 2v1 sub-team interactions before the 1v1 ones, we ensure that the leaf-value estimate reflects the highest degree of team-level synergy available before falling back to independent 1v1 pursuits.
To operationalize the sequential matching process, 
let $ \mathcal{N}_{r}^{\rm rem}\subseteq\mathcal{N}_r$ and $\mathcal{N}_b^{\rm rem}\subseteq\mathcal{N}_b$ denote the set of 
Red and Blue team agents that remain in the game at node $n'$, with $|\mathcal{N}_r^{\rm rem}|=n_r$ and $|\mathcal{N}_b^{\rm rem}| = n_b$. 
We first determine the target number of $2\w1$ and $1\w1$ sub-team 
interactions ($n_{2\w1}$ and $n_{1\w1}$, respectively)  to be considered by checking the partition constraints
\begin{align*}
n_r = 2n_{2\w1} + n_{1\w1}, \quad 
n_b = n_{2\w1} + n_{1\w1}.
\end{align*}
If an exact decomposition is not feasible for a given team configuration, we greedily maximize the number of feasible 2v1 interactions whenever the Red team has numerical superiority; otherwise, only 1v1 interactions are considered. 
We first construct the candidate $2\w1$ interactions by defining the set of unordered Red pairs $\mathcal{P}_{\mathcal{N}_r^{\rm rem}} = \{ \{i,j\} \subseteq \mathcal{N}_r^{\rm rem} \mid i \neq j \}$.
For every pair $\{i,j\} \in \mathcal{P}_{\mathcal{N}_r^{\rm rem}}$ and $k \in \mathcal{N}_b^{\rm rem}$, we populate the $2$v$1$ cost matrix using the $2\w1$ primitive \subgame{} value function : $C_{2\w1}(\{i,j\}, k) := V^*_{2\w1}\left((s^{i}, s^{j}), s^{k}\right).$
We then solve a maximum-sum linear assignment over $(\mathcal{P}_{\mathcal{N}_r^{\rm rem}}, \mathcal{N}_b^{\rm rem})$ using the Hungarian algorithm \cite{hungarian}, yielding a matching $M_{2\w1} \subseteq \mathcal{P}_{\mathcal{N}_r^{\rm rem}} \times \mathcal{N}_b^{\rm rem}$ 
\footnote{Strictly enforcing the disjointness constraint requires solving the combinatorial problem. To enable efficient online planning, we relax this constraint during the $2$v$1$ matching stage, allowing a Red agent to appear in more than one matched pair.}
%
. The top $n_{2\w1}$ highest-value matched tuples are selected, and their constituent agents are removed from the remaining-agent sets $\mathcal{N}_r^{rem}$ and $\mathcal{N}_b^{\rm rem}$. 
For the agents that remain unmatched after the $2$v$1$ matching, we define a $1$v$1$ cost matrix : $C_{1\w1}(i,j) := V^*_{1\w1}(s^{i}, s^{j})$, 
and solve a second linear assignment problem to obtain the $1$v$1$ matching $M_{1\w1} \subseteq \mathcal{N}_r^{\rm rem} \times \mathcal{N}_b^{\rm rem}$. Finally, the leaf value estimate $V^{\text{leaf}}(s)$ 
is computed as the total aggregated sum of the selected primitive sub-games:
\begin{equation}
    V^{\text{leaf}}(s) = \sum_{(\{i,j\},k) \in M_{2\w1}} V^*_{2\w1}\left((s^{i}, s^{j}), s^{k}\right) + \sum_{(i,j)\in M_{1\w1}} V^*_{1\w1}(s^{i}, s^{j}).
    \label{value-estimate}
\end{equation}
\subsection{Action Selection and Update} 
During the selection phase of each simulation $m \in \{1, 2, \dots, M\}$, the algorithm recursively traverses the tree $\mathcal{T}$ by selecting joint actions until a terminal state, an unexpanded node, or the maximum search depth $D$ is reached. 
When an unexpanded node is encountered, it is expanded, and its child nodes are evaluated using the value estimate described in the previous section. 
The resulting value estimate is then backpropagated along the visited trajectory to update the statistics maintained by the selection operator.
As before, let $n$ denote a node in the tree $\mathcal{T}$ visited during this traversal, and let $s_n = (s_n^r, s_n^b)$ denote the joint state associated with that node. 
Action selection follows the standard SM-MCTS framework \cite{mctssmmove} and is treated as a modular component.
At each visited node $n$, each team $\team \in \{r,b\}$ selects an action $a^\team \in \tilde{\mathcal{A}}^\team(s_n^\team)$ according to its selection rule. 
The resulting joint action pair $a=(a^r,a^b)$ determines the successor node in the search tree.
We consider two commonly used action selection operators: Regret matching (RM) and Decoupled {Upper Confidence bounds applied to Trees} (DUCT). 
These are described next.

\paragraph{\textbf{Regret Matching (RM)}:} 

This selection operator applies regret matching~\cite{regret_matching} to the estimated matrix game at every node $n$ in the tree $\mathcal{T}$ to compute a mixed policy for both teams. 
Let $\hat{\boldsymbol{Q}}(s_n) \in \mathbb{R}^{|\tilde{\mathcal{A}}^r(s^r)| \times |\tilde{\mathcal{A}}^b(s^b)|}$ denote the \textit{empirical} action-value matrix estimated dynamically during the tree search, and let $\hat{Q}(s_n,a^r, a^b)$
represent the estimated value for each joint action $(a^r, a^b) \in \tilde{\mathcal{A}}^r(s_n^r) \times \tilde{\mathcal{A}}^b(s_n^b)$ selected $N(s_n,a^r, a^b)$ times. 
The entries of $\hat{\boldsymbol{Q}}(s_n)$ are initialized using the simulated expansion return (see Algorithm~\ref{treesearch-algo} \texttt{ExpandNode} function) 
to ensure that all joint actions have non-zero initial values. 
Additionally, each team $\team \in \{r,b\}$ maintains cumulative regrets $r^\team(s_n, a^\team)$ for not playing joint action $a^\team$ at state $s_n$, and an average policy $\bar{\sigma}^\team(s_n, a^\team)$,
which accumulates the selection policies over all visits. 
All tracking variables are initialized to 0.
The selection and update procedures are summarized in Algorithm~\ref{alg:rm_operator}.
\vspace{-1cm}
\begin{algorithm}[H]
\footnotesize
\caption{Regret (RM) Selection and Update}
\label{alg:rm_operator}
\begin{spacing}{0.9} 
\begin{algorithmic}[1]
\State \textit{Let $s_n$ be the joint state associated with node $n$, and $\phi_r = +1$, $\phi_b = -1$}
\Function{InitializeNodeStats}{$\mathcal{T}, n,(a^r, a^b), Q^{\text{init}}$}
    \State $\hat{Q}(s_n,a^r, a^b) \gets Q^{\text{init}}$, $N(s_n,a^r, a^b) \gets 1$ \Comment{Initialize $\hat{Q}$ estimates and visits}
    \ForAll{team $\team \in \{r,b\}$}
    \Comment{ Initialize regrets and average policy}
        \State $r^\team(s_n,a^\team) \gets 0$, $\bar{\sigma}^\team(s_n,a^\team) \gets 0$
    \EndFor
\EndFunction


\Function{SelectJointAction}{$\mathcal{T},n,m$}
    \ForAll{team $\team \in \{r, b\}$}
        \State \textit{// policy $\propto$ cumulative positive regret}
        \State $\sigma^\team_m(s_n, a^\team) \gets \dfrac{[r^\team(s_n,a^\team)]^+}{\sum_{\tilde{a}^\team}[r^\team(s_n,\tilde{a}^\team)]^+}$ \quad if denominator $> 0$, else uniform 
        \State Sample $a^\team \sim (1-\gamma_{rm})\,\sigma^\team_m(s_n,a^\team) + \dfrac{\gamma_{rm}} {|\tilde{\mathcal{A}}^\team(s_n^\team)|}$ \Comment{$\gamma_{rm}$-greedy exploration}
    \EndFor
    \State \Return $(a^r, a^b)$
\EndFunction

\Function{UpdateNodeStats}{$\mathcal{T},n,(a^r,a^b),v,m$}
    \State \textit{// Update estimated value matrix }
    \State $\hat{Q}(s_n,a^r,a^b) \gets \hat{Q}(s_n,a^r,a^b) + v$, \quad $N(s_n,a^r,a^b) \gets N(s_n,a^r,a^b) + 1$
    
    \State \textit{// Update counterfactual regrets based on sampled opponent action}
    \ForAll{team $\team \in \{r,b\}$, \; $\tilde{a}^\team \in \tilde{\mathcal{A}}^\team(s_n^\team) \setminus \{a^\team\}$}
        \State $r^\team(s_n,\tilde{a}^\team) \gets r^\team(s_n,\tilde{a}^\team) + \phi_\team\!\left(\frac{\hat{Q}(s_n,\tilde{a}^\team, a^{-\team})}{N(s_n,\tilde{a}^\team, a^{-\team})} - v\right)$
    \EndFor
    \State $\bar{\sigma}^\team(s_n,a^\team) \gets \bar{\sigma}^\team(s_n,a^\team) + \sigma^\team_m(s_n,a^\team)$ \Comment{Accumulate average policy}
\EndFunction
\Function{ExtractRootPolicies}{$\mathcal{T},n_0$}
    \ForAll{team $\team \in \{r,b\}$}
        \State $\pi^\team(a^\team) \gets \bar{\sigma}^\team(s_{n_{0}},a^\team) \;/\; \textstyle\sum_{\tilde{a}^\team}\bar{\sigma}^\team(s_{n_0},\tilde{a}^\team)$ \Comment{Normalize accumulated average}
    \EndFor
    \State \Return $(\pi^r, \pi^b)$
\EndFunction
\end{algorithmic}
\end{spacing}
\end{algorithm}
\vspace{-1cm}
\paragraph{\textbf{Decoupled UCT (DUCT):}} 

This selection operator decouples the action-selection at every node $n$ in the tree $\mathcal{T}$ by independently applying the Upper Confidence Bound (UCB) criterion to each team's marginal action-value estimates, thereby ignoring the joint coupling during selection. For each node $n$,
 team maintains marginal action-value estimates $\hat{Q}^\team(s_n,a^\team)$ and visit counts $N^\team(s_n,a^\team)$ for every action $a^\team \in \tilde{\mathcal{A}}^\team(s_n^\team)$,
 along with the total node visit count
$N(s_n)$.
The marginal action values are obtained by aggregating the joint-action value matrix $\hat{\boldsymbol{Q}}(s_n)$ over the opponent's actions. 
Similar to RM, these marginal values are initialized using the expansion return.
The selection and update procedures are summarized in Algorithm~\ref{alg:duct_operator}.

\vspace{-0.5cm}
\begin{algorithm}[h]
\footnotesize
\caption{Decoupled UCT (DUCT) Selection and Update}
\label{alg:duct_operator}
\begin{spacing}{0.8} 
\begin{algorithmic}[1]
\State \textit{Let $s_n$ be the joint state associated with node $n$, $\phi_r = +1$, $\phi_b = -1$}

\Function{InitializeNodeStats}{$\mathcal{T}, n,(a^r,a^b), Q^{\text{init}}$}
    \State $N(s_n) \gets 1$
    \ForAll{team $\team \in \{r, b\}$}
        \State $\hat{Q}^\team(s_n,a^\team) \gets Q^{\text{init}}$, $N^\team(s_n,a^\team) \gets 1$
    \EndFor
\EndFunction
\Function{UpdateNodeStats}{$\mathcal{T},n,(a^r,a^b), v,m$}
    \State $N(s_n) \gets N(s_n) + 1$
    \ForAll{team $\team \in \{r, b\}$} \Comment{Update marginal statistics independently}
        \State $\hat{Q}^\team(s_n,a^\team) \gets \hat{Q}^\team(s_n,a^\team) + v$, $N^\team(s_n,a^\team) \gets N^\team(s_n,a^\team) + 1$
    \EndFor
\EndFunction
\algstore{pgtsbreak}
\end{algorithmic}
\end{spacing}
\end{algorithm}


\begin{algorithm}[t!]
\begin{spacing}{0.8}
\small
\begin{algorithmic}[1]
\algrestore{pgtsbreak}
\Function{SelectJointAction}{$\mathcal{T},n,m$}
    \State \textit{// Red Team (Maximizer) uses UCB, Blue Team (Minimizer) uses LCB}
    \ForAll{team $\team \in \{r, b\}$}
        \State $a^\team \gets \arg\max_{a^\team \in \tilde{\mathcal{A}}^\team(s_n^\team)} \left\{ \phi_\team \frac{\hat{Q}^\team(s_n,a^\team)}{N^\team(s_n,a^\team)} + C \sqrt{\frac{\ln(N(s_n))}{N^\team(s_n,a^\team)}} \right\}$
    \EndFor
    \State \Return $(a^r, a^b)$
\EndFunction
\Function{ExtractRootPolicies}{$\mathcal{T},n_0$}
    \ForAll{team $\team \in \{r, b\}$}
        \State $\pi^\team \gets \arg\max_{a^\team} \; \phi_\team  \dfrac{\hat{Q}^\team(s_{n_0},a^\team)}{N^\team(s_{n_0},a^\team)}$ \Comment{Greedy w.r.t. mean reward}
    \EndFor
    \State \Return $(\pi^r, \pi^b)$
\EndFunction
\end{algorithmic}
\end{spacing}
\end{algorithm}
\begin{algorithm}[H]
\caption{\texttt{Primitive-Guided Tree Search}} 
\label{treesearch-algo}
\small
\begin{algorithmic}[1]
\State \textbf{Require:} Joint state $s_t$, simulations $M$, max depth $D$, game rules $\Gamma^G$, neighbor param $k_r$, discount $\beta$
\State \textbf{Offline:} Solve primitives once before game begins $\pi^*_{1\w1}, V^*_{1\w1}, \pi^*_{2\w1}, V^*_{2\w1}$ 
\State \textbf{Online:} Construct tree $\mathcal{T}$ with root node $n_0$ corresponding to $s_t$

\For{simulation $m = 1$ to $M$} \Comment{Execute M simulations}
    \State \Call{Simulate}{$\mathcal{T}, n_0, m,0$} \Comment{Start simulation at depth d = 0}
\EndFor
\State $(\pi^r, \pi^b) \gets$ \Call{ExtractRootPolicies}{$\mathcal{T}, n_0$} \Comment{$\text{via} {\rm RM} (Alg.~\ref{alg:rm_operator}) \text{or} {\rm DUCT} (Alg.~\ref{alg:duct_operator})$}

\Function{Simulate}{$\mathcal{T}, n, m,d$}
    \State \textit{Let $s_{n}$ be the joint state of the agents associated with node $n$ }
    
    
    \If{$s_{n}$ is empty} \Return $0$ \Comment{Terminal : No active agent remaining}
        
    \ElsIf{$n$ is not expanded in $\mathcal{T}$}
         \State \Call{ExpandNode}{$\mathcal{T},n$}
        \State $(a^r,a^b) \gets$ \Call{SelectJointAction}{$\mathcal{T},n,m$} \Comment{via RM or DUCT}
        \State $Q^{\text{est}} \gets \Call{GetPriorValue}{n, (a^r,a^b)}$ \Comment{set during \textsc{ExpandNode}}
        \State \Call{UpdateNodeStats}{$\mathcal{T},n,(a^r,a^b),Q^{\text{est}},m$} \Comment{via RM or DUCT}
        \State \Return $Q^{\text{est}}$

    \Else  \Comment{Internal node: select joint action and recurse}
        \State $(a^r,a^b) \gets$ \Call{SelectJointAction}{$\mathcal{T},n,m$} \Comment{via RM or DUCT}
        \State  $n' \gets$ child of $n$ after selecting joint action 
        \If{$d+1 = D$} \Comment{Search limit reached; bypass recursion}
            \State $Q^{\text{est}} \gets \Call{GetPriorValue}{n, (a^r,a^b)}$     \Comment{set during \textsc{ExpandNode}}
        \Else \Comment{Recursive descent}
            \State $v_{\text{child}} \gets \Call{Simulate}{\mathcal{T}, n',m,d+1}$
            \State $Q^{\text{est}} \gets \mathcal{R}(s,a^r,a^b) +\beta v_{child}$ \Comment{Query the immediate reward }
        \EndIf
    
    \State \textit{// Backpropagate statistics}
    \State \Call{UpdateNodeStats}{$\mathcal{T},n,(a^r,a^b),Q^{\text{est}},m$} \Comment{via RM or DUCT}
    \State \Return $Q^{\text{est}}$
    \EndIf

\EndFunction{}
\State \textbf{end function}


\Function{ExpandNode}{$\mathcal{T}, n$}
     \State \textit{Let $s_{n}$ be the joint state of the agents associated with node $n$} 
    \State $\tilde{\mathcal{A}}(s_{n}) \gets \Call{PrimitiveGuidedExpansion}{s_{n}, \pi^*_{1\w1}, \pi_{2\w1}^{*}, k_r}$
    
    \ForAll{joint actions $(a^r,a^b) \in \tilde{\mathcal{A}}(s_{n})$}
        \State Add child node $n'$ to  $\mathcal{T}$  
        \State $s_{n'}, \mathcal{R}(s_n,a^r,a^b) \gets \Call{ApplyGameRules}{\mathcal{P}(s_{n}, (a^r,a^b)), \Gamma^G}$ 
        \State \textit{// Compute value estimate for the newly added node}
        \State $V^{\text{leaf}}(s_{n'}) \gets \Call{LeafValueEstimation}{s_{n'}, V^*_{1\w1}, V^*_{2\w1}}$  \Comment{Eq.~\eqref{value-estimate}}
        \State $Q^{\text{init}}(s_n,a^r,a^b) \gets \mathcal{R}(s_n,a^r,a^b)+\beta V^{\text{leaf}}(s_{n'})$
        \State \Call{InitializeNodeStats}{$n, (a^r,a^b), Q^{\text{init}}$} \Comment{Warm-start priors for selection}
    \EndFor
\EndFunction
\State \textbf{end function}
\end{algorithmic}
\end{algorithm} 
\section{Experiments}
\label{sec:experiment}

In this section, we evaluate the performance of the primitive-guided tree search \algacro{} for a graph-based multi-agent Pursuit-Evasion game, in which a team of Pursuers (Red) is tasked with capturing a team of Evaders (Blue) before they can reach the exit nodes (Green) and escape (Fig.~\ref{graphs for the evnironment}).

\subsection{Setup}

To provide a rigorous benchmark, we utilize the open-source \textit{GraphChase} platform \cite{graphchase}, which contains state-of-the-art (SOTA) algorithms for graph-based PEGs. 
While the existing methods on this platform are constrained to $N$-Pursuer versus $1$-Evader $N\text{v}1$ scenarios, our proposed tree search is generalizable to arbitrary multi-agent configuration $N\text{v}M$. 
For a direct and fair comparison with the SOTA baselines, we mainly focus on analyzing the $N\text{v}1$ setting. 

\noindent\textbf{Game rules $\Gamma^G$}: The capture and escape radius are both zero. 
That is, an Evader is captured when a Pursuer occupies the same node, and successfully evades when it reaches an Evasion node.
The game terminates upon the Evader's capture or escape, or when the maximum time horizon $T$ is reached%
\footnote{The time horizon of the game is finite; for simplicity, we solve for the primitive-\subgame{}s by assuming the time horizon is infinity.}. 
If the Evader is neither captured nor has escaped by the end of the time horizon, the Pursuer team receives a reward of +1 for successfully defending.

\begin{figure}[t]
    \centering
    \includegraphics[width=1.0\linewidth]{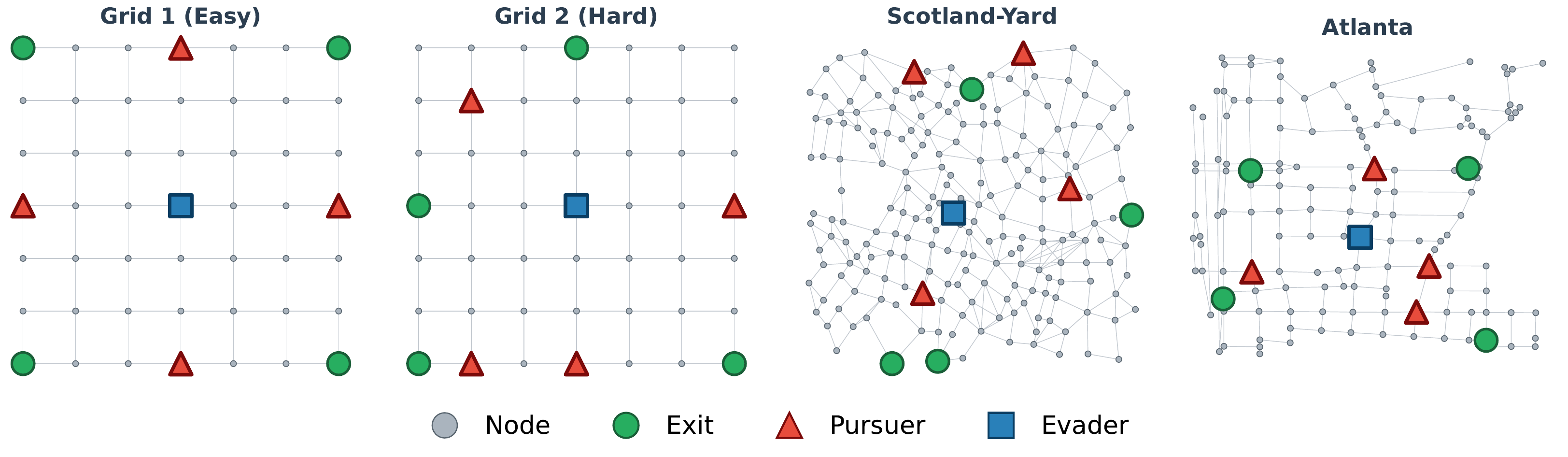}
    \vspace{-0.8cm}
    \caption{Experimental environments used in evaluation.}
      \label{graphs for the evnironment}
\vspace{-0.5cm}
\end{figure}

\noindent\textbf{Evaluation maps:} 
%
We performed experiments on three graph environments natively implemented in GraphChase~\cite{graphchase}: two $7\times7$ grid graphs and a 200-node Scotland Yard graph from GraphChase.
We also added a real-world graph of Atlanta with 151 nodes. 
Details on the graphs and agent configurations are provided in Fig.~\ref{graphs for the evnironment}.

\noindent\textbf{Evaluation Metrics:}
Following the convention established in \textit{GraphChase}, we evaluate the Pursuer team's performance using the following metrics:

\begin{itemize}
\item \textbf{Worst-case utility (WCU):}
Given the initial joint state $s_0=(s_0^p,s_0^e)$, let $\mathcal{T}_e^{T}(s_0^e)$
denote the set of all feasible Evader trajectories from the initial position $s_0^e$ leading to any exit node within the time horizon $T$. 
For a given Pursuer team policy $\pi_p$, 
the worst-case utility is defined as
\begin{equation}
    U_{\text{worst}}(\pi_p\mid s_0,\tau_e) = \min_{\tau_e \in \mathcal{T}_e^{T}(s_0^e)} \mathbb{E}_{\space \pi_p} \left[ \sum_{t=0}^T\mathcal{R}(s_t,a^p_t,a^e_t) \;\middle|\; s_0, \tau_e\right],
    \label{eq:worst_case_utility}
\end{equation}
where the expectation is taken over the Pursuer team’s actions induced by policy $\pi_p$, while the Evader’s action $a_t^e$ deterministically follows the given trajectory $\tau_e$.
This metric evaluates the robustness of a Pursuer team's policy against different Evader behaviors.

\item \textbf{Worst-case utility against shortest-path Evader (SP-WCU):}
To facilitate a fair comparison with the learning-based baselines in GraphChase, which are trained against shortest-path Evaders, we additionally evaluate the worst-case utility over the set of shortest paths connecting the Evader's initial position to each exit node to ensure the training has been completed. 
\item \textbf{Expected reward (ER):}
The expected game reward obtained under the interaction of a particular Pursuer and Evader team policy. 
\end{itemize}
\noindent\textbf{Baselines:} We compared \algacro{} against four benchmarking algorithms spanning learning-based, decomposition-based, and heuristic-based Pursuer policies.

\begin{enumerate}[label=(\roman*)]

    \item \textbf{Multi-Task PSRO (MT-PSRO) \cite{pretrain_PSRO}:} A state-of-the-art PSRO-based framework that solves multi-agent games by iteratively adding new best-response policies to a growing meta-game. It uses Multi-Agent Proximal Policy Optimization (MAPPO) \cite{mappo} as the best-response oracle for the Pursuer team and uses shortest path Evader in the training pipeline. MT-PSRO is natively implemented in the GraphChase package.
    \item \textbf{NSGZero \cite{NSGZero}:} A neural-network-guided 
    Monte-Carlo Tree Search (MCTS) framework developed for network security games. NSGZero is natively implemented in the GraphChase package.
    
    \item \textbf{Decomposition}\footnote{To decompose the Nv1 game into primitive subgames while including all Pursuers, each subgame is defined against the same Evader} \textbf{:}
    The sub-team decomposition policy from Section~\ref{sec:game-decomp}. 
    \item \textbf{Intercepting:} A heuristic rule-based baseline where Pursuers execute a greedy shortest-path policy to intercept the Evader based on its projected path to the nearest exit node.
\end{enumerate}
\vspace{-0.4cm}
\subsection{Results and Discussions}
We present a comprehensive analysis of the experimental results across different graph topologies. 
Table~\ref{baseline_comparison} compares the Pursuer team policy in terms of worst-case utility (WCU), shortest-path worst-case utility (SP-WCU), and expected reward (ER). The expected reward is evaluated over Evader trajectories in $\mathcal{T}_e^T$. To account for stochasticity in the policies, all reported metrics are averaged over $400$ repeated runs for the grid graphs and $200$ runs for the Scotland-Yard and Atlanta graphs.
The tree-search variants use Regret Matching (RM) or Decoupled UCT (DUCT) during action selection\footnote{To ensure a fair comparison, the hyperparameters for both variants are tuned against the shortest-path Evader and held fixed during the worst-case utility evaluation. Additional details are provided in Appendix~\ref{hyperparameters}.}. 

%
Both the proposed \algacro{}-RM and \algacro{}-DUCT consistently achieve strong performance across all graph topologies. 
While Grid 1 serves as a sanity check in which most methods perform well, the performance gap widens as the game complexity increases.
On Grid 2, the tree-search methods maintain high utility under both evaluations, exhibiting clear advantages over the baseline approaches.
The benefit of online search is most apparent for the Scotland-Yard and Atlanta graphs, where the tree-search variants worst-case utility is  substantially higher than the baselines.

\begin{table}[t!]
\centering
\scriptsize
\resizebox{0.9\columnwidth}{!}{%
\begin{tabular}{ll|cccccc}
\toprule
& & \textbf{Intercepting} & \textbf{Decomp.} & \textbf{MT-PSRO} & \textbf{NSGZero} & \textbf{PGTS-RM} & \textbf{PGTS-DUCT} \\
\midrule
\multirow{2}{*}{\textbf{Grid 1 (easy)}}
                 & WCU     & 0.43  & 0.97  & 0.94 & 0.96 & \textbf{0.99}  & \textbf{0.99} \\
                 & ER &0.60  &\textbf{1.0} & 0.97  &\textbf{ 1.0} & \textbf{1.0} & \textbf{1.0}   \\
\midrule
\multirow{3}{*}{\textbf{Grid 2 (hard)}}
                 & SP-WCU & 0.09  & 0.09  & \textbf{0.47} & 0.45 & 0.46 & 0.40   \\
                 & WCU & 0.09  & 0.06  & 0.25   & 0.37 & \textbf{0.46}  & 0.40   \\
                 & ER & 0.78 &0.82 & 0.70  & 0.80 & \textbf{0.90} &  \textbf{0.91} \\
\midrule
\multirow{3}{*}{\textbf{Scotland-Yard}}
                 & SP-WCU  & 0.07  & 0.33  & 0.97   & \textbf{0.98} & 0.82  & 0.96   \\
                 & WCU & 0.00 & 0.27 & 0.05   & 0.00 & \textbf{0.73} & \textbf{0.68}   \\
                 & ER & 0.69 &0.82 & 0.49  & 0.54 & \textbf{0.98} &  \textbf{0.99}  \\
\midrule
\multirow{3}{*}{\textbf{Atlanta}}
                & SP-WCU  & 0.20  & 0.50  & 0.91   & \textbf{1.0} & 0.89  & 0.94   \\
                 & WCU  & 0.00 & 0.47  & 0.49   & 0.00 & 0.87  & \textbf{0.94}   \\
                 & ER & 0.80 & 0.92 & 0.91  &0.94  & \textbf{1.0} & \textbf{1.0}   \\
\bottomrule
\end{tabular}%
}
\vspace{-0.25cm}
\caption{ 
Comparison of the worst-case utility (WCU), shortest-path worst-case utility (SP-WCU), and expected reward (ER) achieved by the Pursuer team across different graph topologies. 
WCU is evaluated against all admissible Evader trajectories in $\mathcal{T}_e^T$, SP-WCU against a shortest-path Evader, and ER against Evader trajectories sampled from $\mathcal{T}_e^T$. The game horizon is T=6 for the grid environments and T=9 for the Scotland Yard and Atlanta environments.}
\label{baseline_comparison}
\vspace{-0.5cm}

\end{table}
The sub-team decomposition framework also provides a valuable strategic prior. The Decomposition Pursuer remains competitive with learned baselines against a  strategic Evader. 
However, decomposition alone is much less effective than tree search across all graph topologies, underscoring the importance of team coordination. 
This limitation arises because the decomposition-based approach focuses on agent coordination within each primitive \subgame{}, but ignores joint coordination among agents at the team level.
To further evaluate performance, we compared the various Pursuer team policies 
against \algacro{} Evader in Table \ref{tree_search_Evader_comparison}. 
Our \algacro{} Pursuers consistently achieve the highest rewards. 

In addition, although MT-PSRO and NSGZero achieve stronger performance against shortest-path Evaders, their policies appear to be overfitted to the shortest-path Evaders (SP-WCU) used during training.
As a result, they struggle against more complex or unseen Evader behaviors, leading to substantially lower worst-case utility (Table~\ref{baseline_comparison}), as well as weaker performance against PGTS-based Evaders (Table~\ref{tree_search_Evader_comparison}).
In contrast, the \algacro{} Pursuer maintains robust performance against all variants of Evader policies since it evaluates the game tree dynamically, rather than relying solely on precomputed historical data distributions.
\begin{table}[!t]
\centering
\scriptsize
\resizebox{0.85\columnwidth}{!}{%
\begin{tabular}{lll|cccccc}
\toprule
\multicolumn{3}{c|}{%
\diagbox[width=18em,height=2em]
{\textbf{Evader policies}}
{\textbf{Pursuer policies}}
}& \textbf{Intercepting} & \textbf{Decomp.} & \textbf{MT-PSRO} & \textbf{NSGZero} & \textbf{PGTS-RM} & \textbf{PGTS-DUCT} \\
\midrule
\multirow{4}{*}{\textbf{Grid 1 (easy)}}
& \multirow{2}{*}{\textbf{PGTS-RM}}   & T=6  & 0.61 & \textbf{1.0}  & 0.99 & \textbf{1.0}   & \textbf{1.0}    & \textbf{1.0}   \\
&                                      & T=12 & 0.10   & \textbf{1.0}  & 0.62  & 0.14 & \textbf{0.98}  & \textbf{0.98} \\
\cmidrule{2-9}
& \multirow{2}{*}{\textbf{PGTS-DUCT}} & T=6  & 0.50  & \textbf{1.0}  & 0.98  & \textbf{1.0 }  & \textbf{1.0 }   & \textbf{1.0}   \\
&                                      & T=12 & 0.08  & \textbf{1.0}  & 0.62   & 0.06  & \textbf{0.98} & \textbf{1.0} \\
\midrule
\multirow{4}{*}{\textbf{Grid 2 (hard)}}
& \multirow{2}{*}{\textbf{PGTS-RM}}   & T=6  & 0.09   & 0.19 & 0.03   & 0.04  & 0.23   & \textbf{0.27}  \\
&                                      & T=12 & 0.02 & 0.15 & 0.01 & 0.02  & 0.23   & \textbf{0.27}  \\
\cmidrule{2-9}
& \multirow{2}{*}{\textbf{PGTS-DUCT}} & T=6  & 0.03   & \textbf{0.15} & 0.02 & 0.07  & 0.07   & 0.11  \\
&                                      & T=12 & 0.02   & \textbf{0.11} & 0.01   & 0.04  & 0.07   & \textbf{0.10}  \\
\midrule
\multirow{4}{*}{\textbf{Scotland-Yard}}
& \multirow{2}{*}{\textbf{PGTS-RM}}   & T=9 &  0.02  & 0.93 & 0.76 & 0.77 & \textbf{0.98}& \textbf{0.99} \\
&                                      & T=20 &  0.01  & 0.80 & 0.14 & 0.18 &0.87 & \textbf{0.95} \\
\cmidrule{2-9}
& \multirow{2}{*}{\textbf{PGTS-DUCT}} & T=9  & 0.0  & 0.94 & 0.72 & 0.13 & \textbf{0.98}  &  \textbf{0.99} \\
&                                      & T=20 &  0.0  &  0.72 & 0.18 & 0.03 &  \textbf{0.94}& \textbf{0.96} \\
\midrule
\multirow{4}{*}{\textbf{Atlanta}}
& \multirow{2}{*}{\textbf{PGTS-RM}}  & T=9  &  0.21  & 0.87   & 0.87 &  \textbf{1.0} & \textbf{1.0} & \textbf{1.0} \\
&                                      & T=20 &  0.15  & 0.71 & 0.30 &  0.31 & 0.97 &  \textbf{0.99 }\\
\cmidrule{2-9}
& \multirow{2}{*}{\textbf{PGTS-DUCT}} & T=9  &  0.20  & 0.82   & 0.97 & \textbf{1.0 } & \textbf{1.0} &  \textbf{1.0}\\
&                                      & T=20 &  0.14  & 0.87 & 0.31 & 0.25  & 0.93 &  \textbf{1.0 }\\
\bottomrule
\end{tabular}%
}
\vspace{-0.25cm}
\caption{Comparison of the expected Pursuer team reward against tree search Evader across two time horizons $T$. PGTS-RM and PGTS-DUCT denote the proposed tree search with Regret Matching and Decoupled UCT selection operators, respectively.}
\label{tree_search_Evader_comparison}
\end{table}

\subsection{Ablation Study}

To better understand the contribution of the individual components of the proposed \algacro{} framework, we performed an ablation study on both the Regret Matching (RM) and Decoupled UCT (DUCT) variants.
Specifically, we evaluated the importance of (i) primitive-guided expansion to focus the search on strategically relevant actions, and (ii) higher-order $2\w1$ primitive interactions to capture coordinated team behavior. 
We compared the full \algacro{} against two degraded variants.
\vspace{-0.2cm}
\begin{enumerate}
\item \textbf{Tree search w/o Primitive Guided Expansion:} Primitive Nash \subgame{} policies are removed from the expansion and rollout phases and replaced with hand-crafted heuristic policies. 
The Evader selects shortest-path actions toward all the exit nodes, while the Pursuers select shortest-path actions toward the Evader, the nearest exit node, and an interception point. The full decomposition-based leaf-value estimator, including both $1\w1$ and $2\w1$ primitives, is retained.

\item 
\textbf{PGTS with only one primitive 1v1:} The tree-search algorithm with only the $1\w1$ primitive subgame. 
Specifically, the higher-order $2\w1$ primitive is omitted from both the guided expansion phase and the leaf-value estimator.

\end{enumerate}
\vspace{-0.0cm}
The results of the ablation study are summarized in Table~\ref{ablation_study}. Overall, both primitive-guided expansion and higher-order $2\w1$ primitives contribute meaningfully to the performance of \algacro{}.
\begin{table}[t]
\centering
\scriptsize
\resizebox{0.85\columnwidth}{!}{%
\begin{tabular}{l|cccc}
\toprule
 & \textbf{Grid 1 (easy)} & \textbf{Grid 2 (hard)} & \textbf{Scotland-Yard} & \textbf{Atlanta} \\
\midrule
\textbf{PGTS-RM}           & \textbf{0.99} & \textbf{0.46}   & \textbf{0.73}  & \textbf{0.89}  \\
\textbf{TS-RM-wo-PG-Expansion}     & 0.89 &0.27  &  0.58 & 0.73  \\
\textbf{PGTS-RM-1v1} & 0.65 & \textbf{0.47} & 0.15 & 0.68 \\
\midrule
\textbf{PGTS-DUCT}           & \textbf{0.99} & \textbf{0.40}    & \textbf{0.68}  & \textbf{0.94}  \\
\textbf{TS-DUCT-wo-PG-Expansion}     & 0.95 & 0.22  & 0.30 & 0.90 \\
\textbf{PGTS-DUCT-1v1} & 0.81 &  0.31 & 0.12 & 0.69 \\
\bottomrule
\end{tabular}%
}
\vspace{-0.25cm}
\caption{Comparison of the worst-case utility (WCU) for the variants of the tree search Pursuer team policy across different graphs. The results isolate the impact of primitive-guided expansion and $2$v$1$ primitive inclusion.}
\vspace{-0.25in}

\label{ablation_study}
\end{table}
While primitive-guided expansion improves the quality of exploration, the inclusion of $2$v$1$ primitive allows the search to reason about coordinated team interactions beyond independent $1$v$1$ pursuits. For the Grid 2 configuration, the performance gap is relatively small, because the chosen initial configuration leads to game termination within a few steps and thus does not require deep lookahead.
%
The benefits of both components become increasingly pronounced on larger graph instances, such as Scotland-Yard and Atlanta, where the longer game horizon makes effective exploration and accurate value estimation particularly important.


\section{Conclusion and Future Work}
\label{sec:conclusion}

We have presented Primitive-Guided Tree Search (\algacro{}), a hybrid framework that integrates offline Nash computation with online tree search. By solving a collection of small primitive \subgame{}s offline and embedding their policies and value functions into the search process, \algacro{} balances computational tractability with coordinated team-level decision making. \algacro{} achieves robust performance across diverse graph topologies against state-of-the-art methods.

Future work will focus on extending the proposed framework beyond pursuit-evasion settings to a broader class of multi-agent games. We also plan to evaluate the approach across a wider range of team sizes and agent configurations to better understand its scalability and robustness. 
Finally, the primitive-guided tree-search framework is largely agnostic to the underlying information structure of the game. 
It is therefore worth investigating extensions to settings with partial observability and asymmetric information, thus enabling the approach to address a broader range of realistic multi-agent settings.

\noindent\textbf{Acknowledgements:} This research was supported by the U.S. Army Research Laboratory through the Distributed and Collaborative Intelligent Systems and Technology (DCIST) Collaborative Research Alliance under Cooperative
Agreement No. W$911$NF-$17$-$2$-$0181$.

\noindent\textbf{Disclosure of Interests:} The authors have no competing interests to declare that are relevant to the content of this article.

\appendix
\section{Appendix} \label{hyperparameters}

All experiments were performed on a workstation equipped with a $24$-core 5.7GHz Intel Core Ultra $9$ $285$K CPU and an NVIDIA RTX $5090$ GPU. 
For the offline computation of the primitive \subgame{} solutions, we utilized the Nash-equilibrium solver developed in \cite{guan2021learning}. 
Since the proposed framework relies on pre-computed primitive policies and value functions, the computational cost of solving these primitives is an important consideration. Table~\ref{tab:computation_time} reports both the order of the corresponding state-space sizes and the offline computation times for the $1\w1$ and $2\w1$ primitive \subgame{}s across the different graph topologies.

\begin{table}[!t]
\centering
\scriptsize
\resizebox{0.85\columnwidth}{!}{%
\begin{tabular}{l|c|c|c|c|c|c}
\toprule
 & \textbf{No. of Nodes} & $\mathbf{S_{1\w1}}$ & $\mathbf{S_{2\w1}}$ & $\mathbf{S}$ & \textbf{$1\w1$ Primitive} & \textbf{$2\w1$ Primitive} \\
\midrule
\textbf{Grid 1 (easy)}   & 49  & $2.4 \times 10^{3}$ & $6.0 \times 10^{4}$ & $1.3 \times 10^{7}$ & 0.9\,s  & 56.0\,s \\
\textbf{Grid 2 (hard)}   & 49  & $2.4 \times 10^{3}$ & $6.0 \times 10^{4}$ & $1.3 \times 10^{7}$ & 0.9\,s  & 65.3\,s \\
\textbf{Atlanta}         & 151 & $2.3 \times 10^{4}$ & $1.7 \times 10^{6}$ & $3.4 \times 10^{9}$  & 22.1\,s     & 2948.7\,s     \\
\textbf{Scotland-Yard}   & 200 & $4.0 \times 10^{4}$ & $4.0 \times 10^{6}$ & $1.4 \times 10^{10}$ & 33\,s    & 23500.7\,s      \\
\bottomrule
\end{tabular}%
}
\vspace{-0.25cm}
\caption{Order of State-space sizes and offline computation times for the primitive \subgame{}s across the evaluated graph topologies.}
\label{tab:computation_time}
\vspace{-0.7cm}
\end{table}
For online planning, we limit the tree-search horizon to a maximum depth of four and perform at most $2,000$ simulations at each decision step. Among the four phases of the tree-search procedure, the expansion phase is the most computationally expensive, as it requires generating candidate actions and estimating value.
To improve efficiency, these computations are vectorized, resulting in an average decision time of approximately $0.5$ seconds per planning step.
The hyperparameters are summarized in Table~\ref{tab:hyperparams}. Across all environments, we use a discount factor of $\beta=0.95$.
\begin{table}[H]
\vspace{-0.55cm}
\centering
\scriptsize
\resizebox{0.75\columnwidth}{!}{%
\begin{tabular}{l|c|c}
\toprule
 & \textbf{PGTS-RM} & \textbf{PGTS-DUCT} \\
\midrule
\textbf{Grid 1 (easy)}    & $M=2000, D=4, \gamma_{rm}=0.6,k_r =4$ & $M=1000, D=3, C=2.0,k_r =4$ \\
\textbf{Grid 2 (hard)}    & $M=2000, D=2, \gamma_{rm}=0.6, k_r =4$ & $M=1000, D=4, C=2.0, k_r =4$ \\
\textbf{Scotland-Yard}    & $M=2000, D=4, \gamma_{rm}=0.4, k_r =3$ & $M=2000, D=4, C=2.0, k_r =3$ \\
\textbf{Atlanta}          & $M=2000, D=4, \gamma_{rm}=0.6, k_r =4$ & $M=2000, D=4, C=2.0, k_r =4$ \\
\bottomrule
\end{tabular}%
}
\vspace{-0.25cm}
\caption{Hyperparameters for the primitive-guided tree search algorithms across different environments. $M$: number of simulations, $D$: search depth, $\gamma_{rm}$: exploration rate (RM), $C$: exploration constant (DUCT), $k_r$: number of neighboring Red agents considered during primitive-guided expansion.}
\label{tab:hyperparams}
\vspace{-0.7cm}
\end{table}

\bibliographystyle{splncs04}
\bibliography{sample}

\end{document}